\documentclass{iaus}
\usepackage{graphicx}

\title[JD 11.~~Gas dynamics in whole galaxies: SPH] 
{Gas dynamics in whole galaxies: SPH}

\author[Clare Dobbs]   
{Clare Dobbs$^{1,2}$
}
\affiliation{$^1$Max-Planck-Institut f\"ur extraterrestrische
  Physik, \\
Giessenbachstra\ss{}e, D-85748 Garching, Germany \\ email: {\tt cdobbs@mpe.mpg.de}
\\[\affilskip]

$^2$Universit\"ats-Sternwarte M\"unchen, Scheinerstra\ss{}e 1, D-81679
M\"unchen, Germany}
\pubyear{2008}
\volume{xxx}  
\pagerange{119--126}
\jname{Title of your IAU Symposium}
\editors{A.C. Editor, B.D. Editor \& C.E. Editor, eds.}
\begin{document}

\maketitle

\begin{abstract}
I review the progress of SPH calculations for modelling galaxies, and
resolving gas dynamics on GMC scales. SPH calculations first
investigated the response of isothermal gas to a spiral potential, in the absence of self
gravity and magnetic fields. Surprisingly though, even these
simple calculations
displayed substructure along the spiral arms. Numerical tests indicate
that this substructure is still present at high resolution (100
million particles, $\sim$ 10 pc), and is independent of the initial particle
distribution. One interpretation of
the formation of substructure is that smaller clouds can agglomerate
into more massive GMCs via dissipative collisions. More recent
calculations have investigated how other processes, such as the
thermodynamics of the ISM, and self gravity affect this simple
picture. Further research has focused on developing models with a more
realistic spiral structure, either by including stars, or incorporating a
tidal interaction.
\end{abstract}

\firstsection 
\section{Introduction}
The last decade has seen massive progress in the use of numerical
calculations to model the ISM, both on galactic and subgalactic scales. 
Early particle simulations (Levinson \& Roberts 1981; Kwan \& Valdes 1983, 1987;
Hausman \& Roberts 1984;
Tomisaka 1984, 1985; Roberts \& Stewart 1987) were largely limited to
modelling individual clouds as ballistic particles. Now, 
hydrodynamic simulations of galaxies are able to resolve
GMCs. Furthermore they can
capture the dynamical evolution of molecular (and atomic) clouds,
as they form, interact and disperse.

To maximise resolution, hydrodynamic simulations frequently
apply an external potential, which includes the dark matter halo and
stellar disc. Numerous calculations have also
included a spiral component to the stellar potential in order to
simulate a grand design galaxy. Surprisingly, calculations which 
model the gas response to a
spiral potential show the presence of substructure in the gas,
even without magnetic fields or self gravity (Wada \& Koda 2004, Dobbs \&
Bonnell 2006). This substructure becomes more clearly visible as 
the spiral arm clumps are sheared out into spurs, or feathers, in the interarm
regions.  Such structure is found to occur in both grid
(Wada \& Koda 2004, Kim \& Ostriker 2006) and SPH (Wada \& Koda 2004,
Dobbs \& Bonnell 2006) codes.

Wada \& Koda 2004 interpreted the formation of substructure along the spiral arms in
terms of Kelvin
Helmholtz instabilities.
However Dobbs et al. 2006 provided a different explanation for the
formation of spiral arm clumps. They likened the passage of gas
through the spiral shock to a queue of traffic, where gas particles
(or clumps) become bunched together. This occurs because the gas
particles undergo dissipative collisions in the spiral arms. This process
resembles prior models of cloud coalescence, but the accretion and dispersal of gas
into and from clumps is more continuous. Also, there is no
need for the gas to be molecular, although for some environments e.g. M51 it
may well be. The main requirement (Dobbs \& Bonnell 2006) is that the gas is cold,
so that the spiral shock is strong, and the gas pressure (which would
cause clouds to diffuse) is minimal. The spacing of the clumps is
$\sim \sigma_v T$ where $\sigma_v$ is the turbulent velocity dispersion of the
gas, and $T$ the time spent in the spiral arms (Dobbs, Bonnell \&
Pringle 2006, Dobbs 2008). In simulations with higher spiral
potentials 
(and consequently higher arm to interarm contrasts), 
$T$ increases and the spacing between the clumps increases (Dobbs 2008). 
\begin{figure}[b]
\begin{center}
\centerline{
 \includegraphics[width=2.8in]{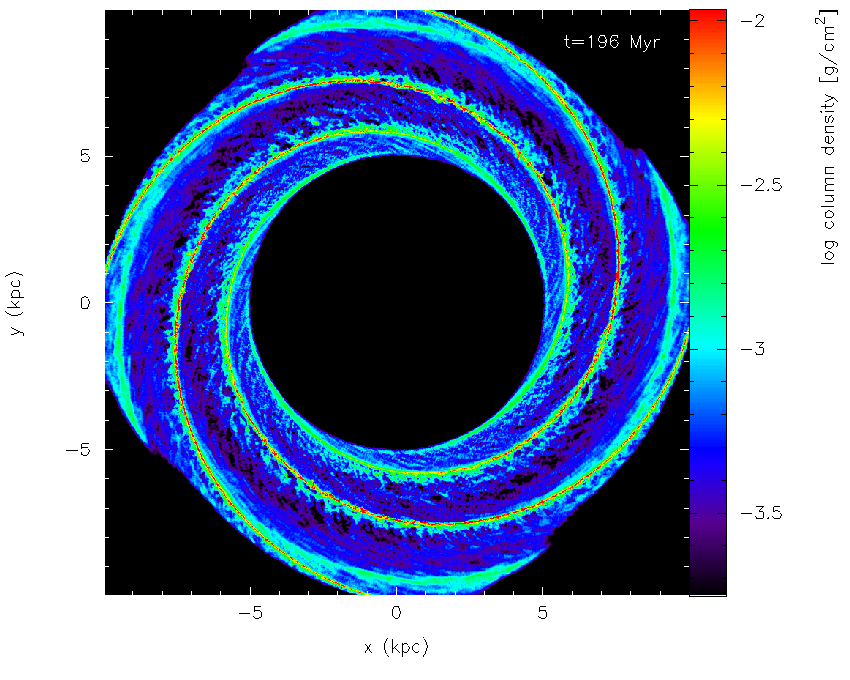} 
 \includegraphics[width=2.8in]{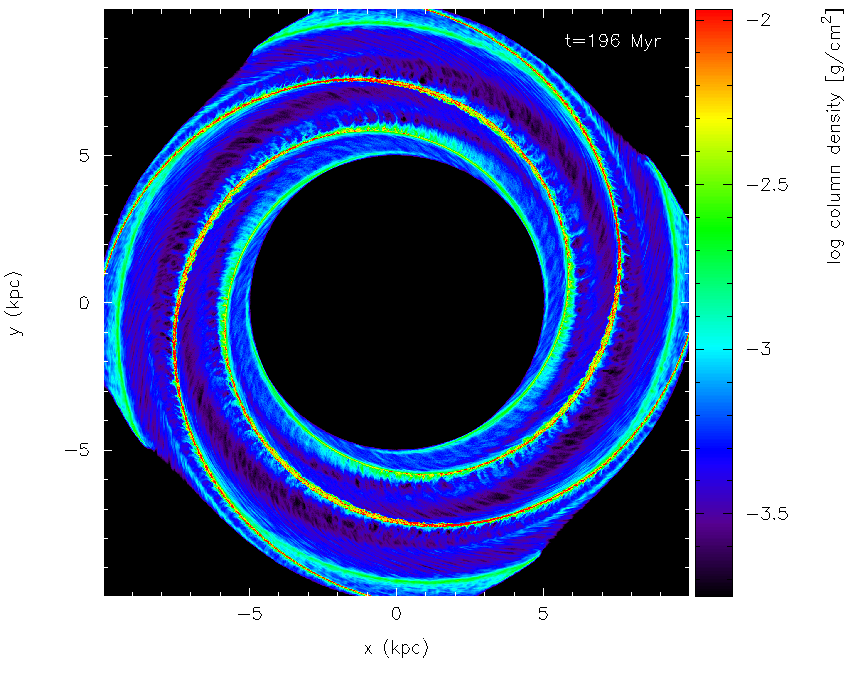}} 
 \includegraphics[width=2.8in]{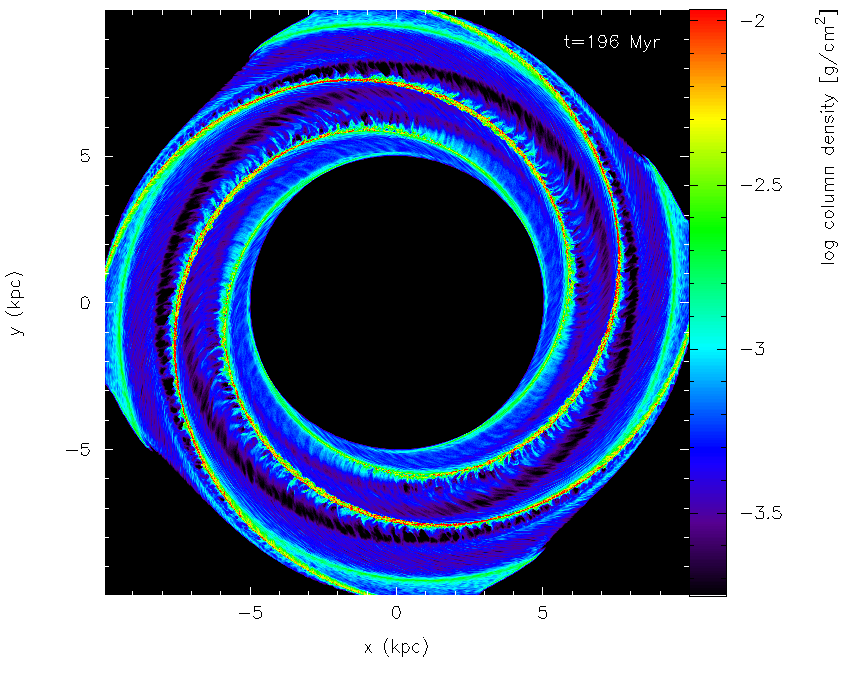} 
 \caption{Resolution test for hydrodynamic calculations with a spiral
   potential alone. The number of particles in each panel is 1 million
 (top left), 20 million (top right) and 100 million (lower).}
   \label{fig1}
\end{center}
\end{figure}
\section{Resolution studies}
A potential issue with the results of these calculations is that the
structure may be resolution dependent, or even disappear at high
resolution. In Fig.~1, we show this is not the case. Similar structure is
apparent with 1, 20 and 100 million particles, the latter
corresponding to a mass resolution of 10 M$_{\odot}$ and a spatial resolution
of $<1$ pc. It is difficult to quantify the spacing between the
features along the spiral arms - partly because there is no exact
periodicity. Nevertheless in Fig.~2, we show the auto-correlation
function, taken from the distribution of particles in $log(r)$ versus
$\theta$ space. The position of the first peak indicates the spacing
which provides the best correlation. We see that the peak is in a
similar location for each resolution, and also that the typical
spacing between subsequent peaks is similar. Unfortunately the spacing
of the clumps has not reached a maximum at this time (200 Myr) as it was 
too time consuming to run the 100 million particle simulation further.

\begin{figure}[t]
\begin{center}
 \includegraphics[width=2.8in]{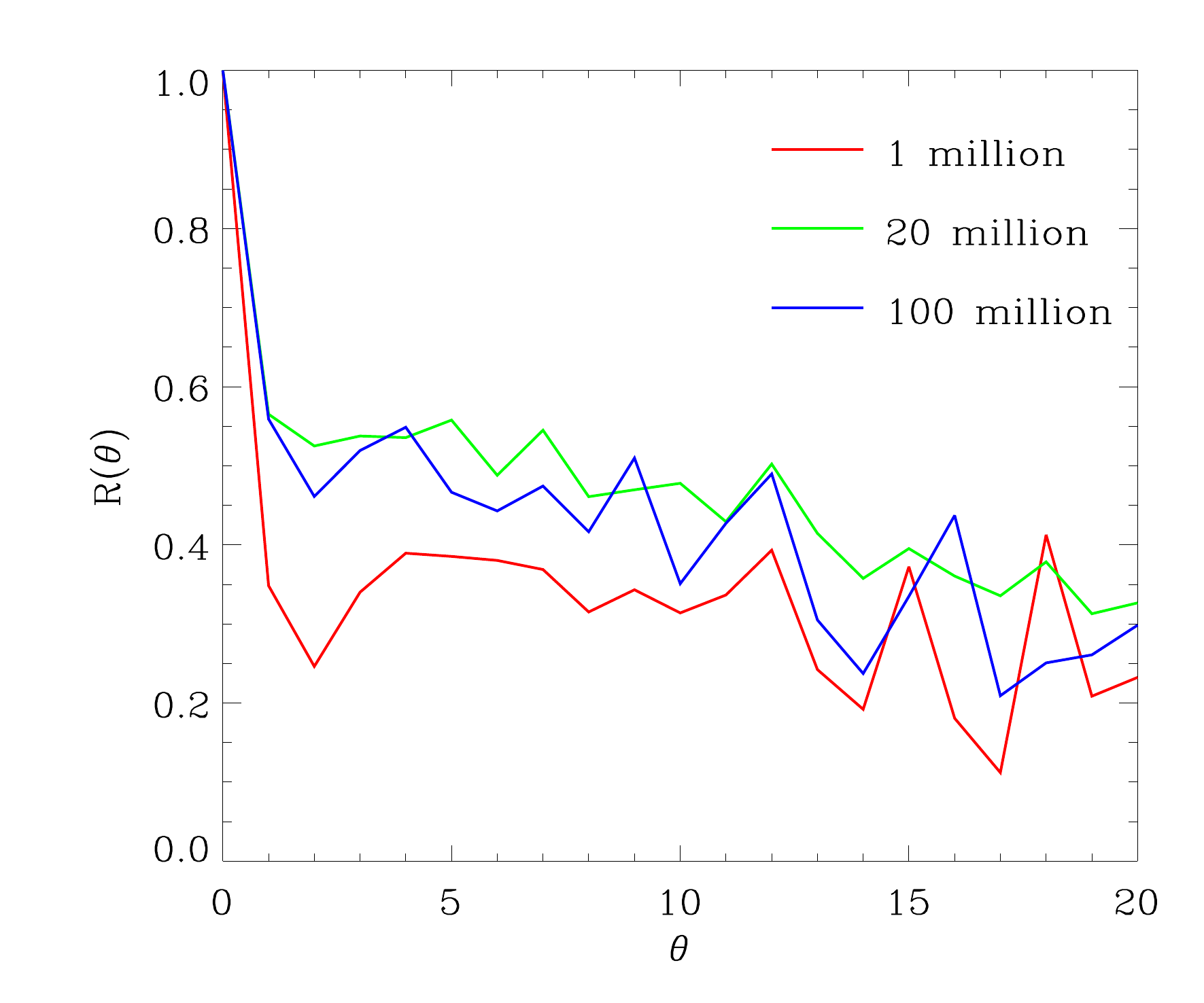}
 \caption{The auto-correlation function ($R$) is plotted for the
   distribution of particles along a spiral arm. $\theta$ is the
   angle, or distance along the spiral arm. The correlation
   function is the sum of the distribution of particles, multiplied by the
   distribution shifted by $\theta$. Thus the first peak represents
   the best-fit separation of the spurs. The similarity in location of
 the first peaks, and spacing of subsequent peaks implies that the
 spacing is not resolution dependent. The spacing is around 200 pc,
 but has not yet reached a maximum (which is closer to 700 pc (Dobbs, Bonnell \&
Pringle 2006)).}
   \label{fig1}
\end{center}
\end{figure}
In Fig.~3, we show further tests, varying the viscosity and the
initial particle distribution. We use 1 million particles in each calculation. The interarm spurs occur in all the
cases, and in particular using an initially uniform
particle distribution (rather than a random distribution, which is
used for all the other calculations) makes negligible difference to the results.  
\begin{figure}[t]
\begin{center}
\centerline{ \includegraphics[width=2.8in]{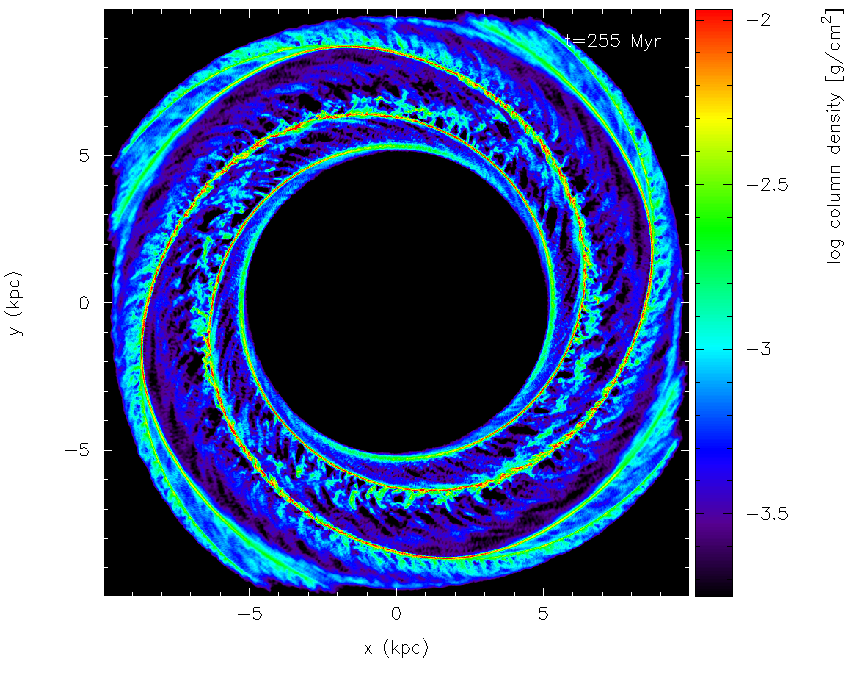}
 \includegraphics[width=2.8in]{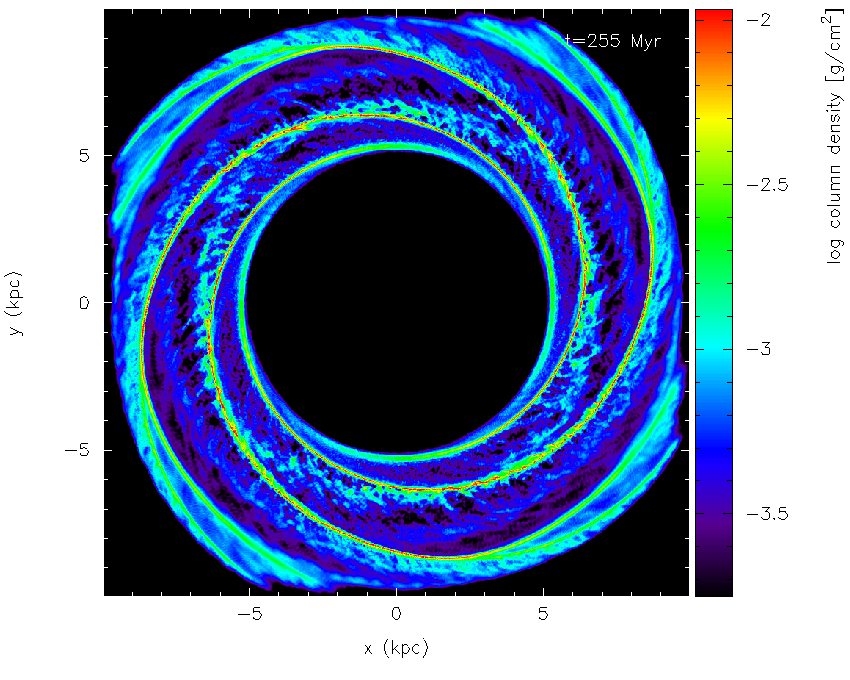}}
\centerline{ \includegraphics[width=2.8in]{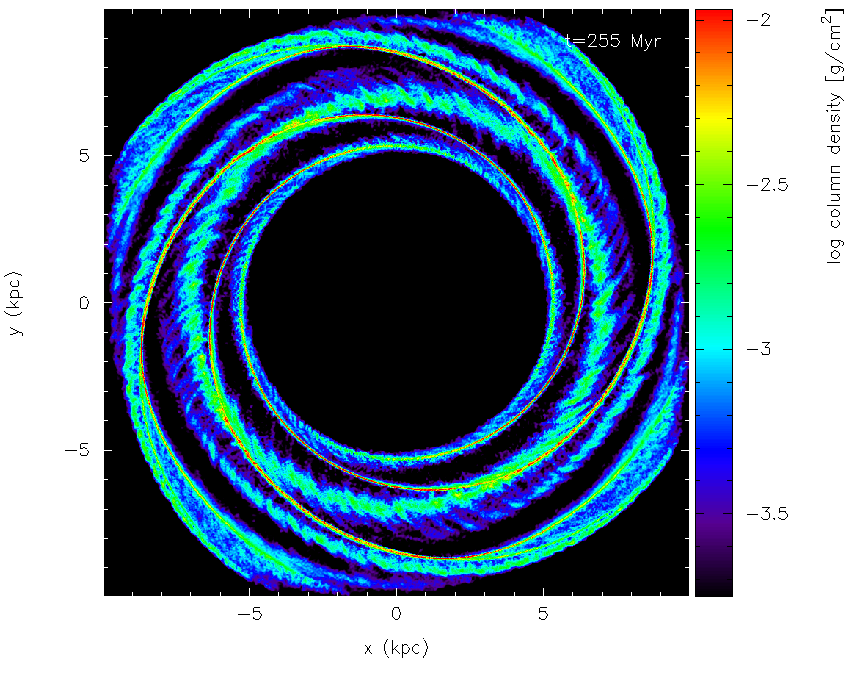}
 \includegraphics[width=2.8in]{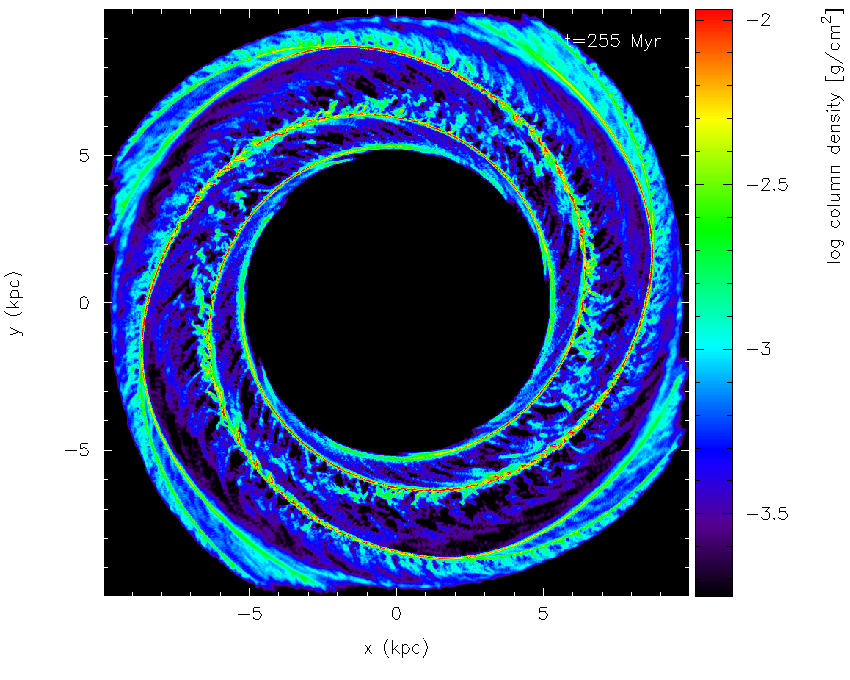}}
 \caption{Column density images from SPH simulations of isothermal gas
   subject to a spiral potential. Three plots show different
   artificial viscosity settings: the
 standard case, $\alpha=1$, $\beta=2$ (top left); $\alpha=2$,
 $\beta=4$ (top right); Balsara switch with $\alpha=1$, $\beta=2$
 (lower left). Finally the lower right panel shows the column density
 when the particles are initially set up uniformly, on concentric
 circles. The difference for the Balsara switch is probably because
 the shock is not captured so well.}
   \label{fig1}
\end{center}
\end{figure}
\section{Thermodynamics and molecular hydrogen formation}
Early calculations adopted an isothermal medium (Slyz et al. 2003,
Wada \& Koda 2004, Dobbs \& Bonnell 2006),
but more recent calculations have included a much more sophisticated
treatment of the ISM (Dobbs et al. 2008, Peluppessy \& Papadopoulos
2009). In Dobbs et al. 2008, we used
the thermodynamics and chemistry of Glover \& MacLow 2007. The results of these
calculations indicate that GMC formation occurs in the same way as
isothermal simulations of cold gas, by the coalescence of clouds in
the spiral shock, as most of the gas (70\%) tends to be cold.

By including H$_2$ formation, we can start to make estimates on cloud
formation timescales, and cloud lifetimes. In Dobbs et al. 2008, we found H$_2$
forms on timescales of Myrs, whilst once formed, H$_2$ lasts 10's of
Myrs. However these calculations do not include any feedback
processes, only H$_2$
photodissociation. Lifetimes based around the cloud dynamics, rather
than H$_2$ fractions, are difficult to determine, since the clouds
evolve dynamically
 on relatively short timescales.

The inclusion of H2, and CO also allows the opportunity of comparison
with observations. Pelupessy \& Papadopoulos 2009 (Fig.~5) calculate synthetic UV and CO maps of their galaxy models. In
Douglas et al. 2010, we present synthetic HI maps, which show  abundant HI self
absorbtion features (also Acreman et al., these proceedings).

\section{Magnetic fields}
Magnetic fields have recently been implemented in simulations of
galaxies (Dobbs \& Price 2008, Dolag \& Stasyszyn 2008, Kotarba et al. 2009)
using the Euler potential method (Price, these
proceedings). Although the current Euler potential implementation is known to have
limitations, in particular amplification of the field due to winding
is not captured, this method does allow a first order study of
the effects of magnetic fields. 

In Dobbs \& Price 2008 we find that the main effect of the magnetic
fields is to contribute pressure to the ISM. The result is similar to
increasing the thermal pressure, as the spiral arms and spurs are
less dense and
more diffuse. However spurs are still visible for plasma $\beta>0.1$.
Kotarba et al. 2009 study the amplification of magnetic fields
in SPH simulations with stars and gas, where the spiral arms arise
consistently. They compare calculations with the Euler method, and
direct calculation of B, finding that even though the amplification is
likely underestimated with the Euler method, the large value of
\textit{div} $B$ when using the direct calculation of $B$ render this method
much less reliable.
\section{Self gravity}
In Dobbs 2008, we perform simulations with self gravity and magnetic
fields. Self gravity enhances the formation of molecular clouds, by
increasing the frequency and success (i.e. the likelihood that clouds
merge rather than get disrupted) of collisions. Self gravity also
leads to 
gravitational instabilities in the gas. The net result is that clouds
reach higher densities compared to the ambient medium, and
are easier to distinguish along the spiral arms. The
difference in structure becomes more noticeable with increasing
surface densities - the gas distribution when $\Sigma=4$ Mpc$^{-2}$ is
similar whether or not self gravity is included, but self gravity has
a much stronger impact on the global structure when
$\Sigma=20$ Mpc$^{-2}$ (Fig.~4). The surface density in the solar neighbourhood
lies in between these regimes (Wolfire et al. 2003).

\section {Feedback}
So far the effect of feedback on molecular cloud formation, and the
evolution of spurs has not been studied for grand design galaxies with
SPH. This has been studied with grid codes, e.g. Wada 2008 find that
supernovae have a secondary effect, and spurs are still evident.
\begin{figure}[t]
\begin{center}
\centerline{ \includegraphics[width=2.8in]{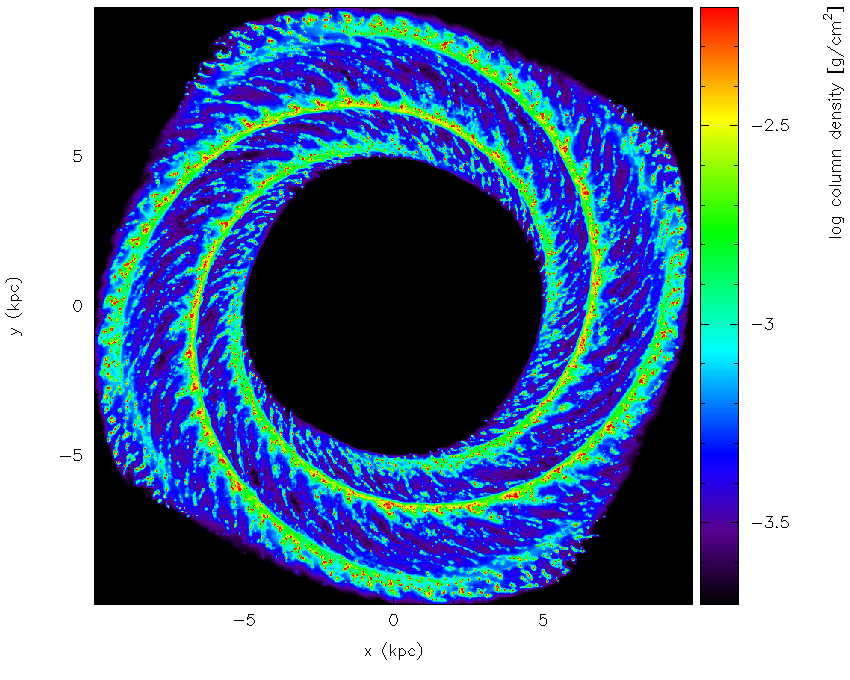}
\includegraphics[width=2.8in]{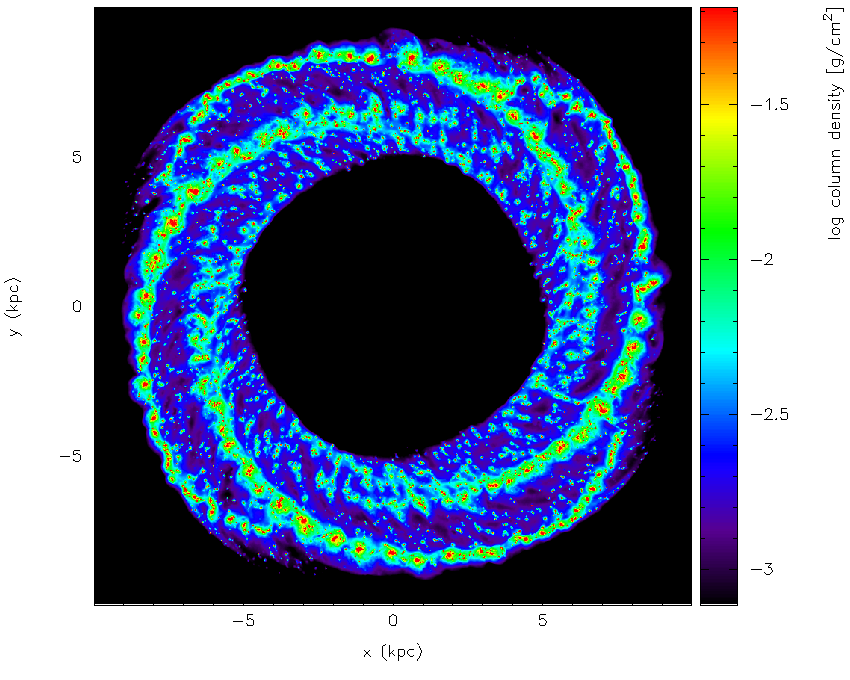}}
 \caption{Column density from simulations with surface densities of 4
   M$_{\odot}$pc$^{-2}$ (left) and 20  M$_{\odot}$pc$^{-2}$
   (right). The simulations include self gravity and magnetic fields,
   adopting an isothermal two-phase medium of cold and warm gas. Self
   gravity has little effect on the global structure in the low
   surface density case, but aids the formation of more massive
   complexes in the higher density case.}
   \label{fig1}
\end{center}
\end{figure}

\section{SPH simulations without a spiral potential}
Several authors have also modelled flocculent galaxies, where the
structure is solely due to gravitational and thermal instabilities,
and is
presented elsewhere in these proceedings (Wada).

Pelupessy \& Papadopoulos (2006,2009) also model flocculent spirals
(Fig.~5), and show that they are able to reproduce characteristics of the ISM such as PDFs and
the pressure H$_2$ relation (Blitz \& Rosolowsky 2004) reasonably well. They also use the
amount of H$_2$ formation to estimate the star formation efficiency,
which is then used in their models of
feedback. One of their main results is that for gas rich, and
metal poor galaxies, they find strong deviations from the star
formation rates in their models and those predicted from the
Schimdt-Kennicutt relation (Pelupessy \& Papadopoulos 2010).

Alternatively a grand design spiral pattern can be obtained more
realistically by modelling an interaction with another galaxy (see
Struck, these proceedings). In Dobbs et al. 2010, we modelled the
grand design spiral M51 (Fig.~6), and
showed that the interaction with the companion can reproduce both the
observed spiral pattern, and detailed features in the gas. We also
found the pattern highly time dependent. Other studies
have concentrated on reproducing the star formation histories of interacting
galaxies, e.g. Karl et al. 2010.
 \begin{figure}[b]
\begin{center}
\centerline{ \includegraphics[width=4.5in]{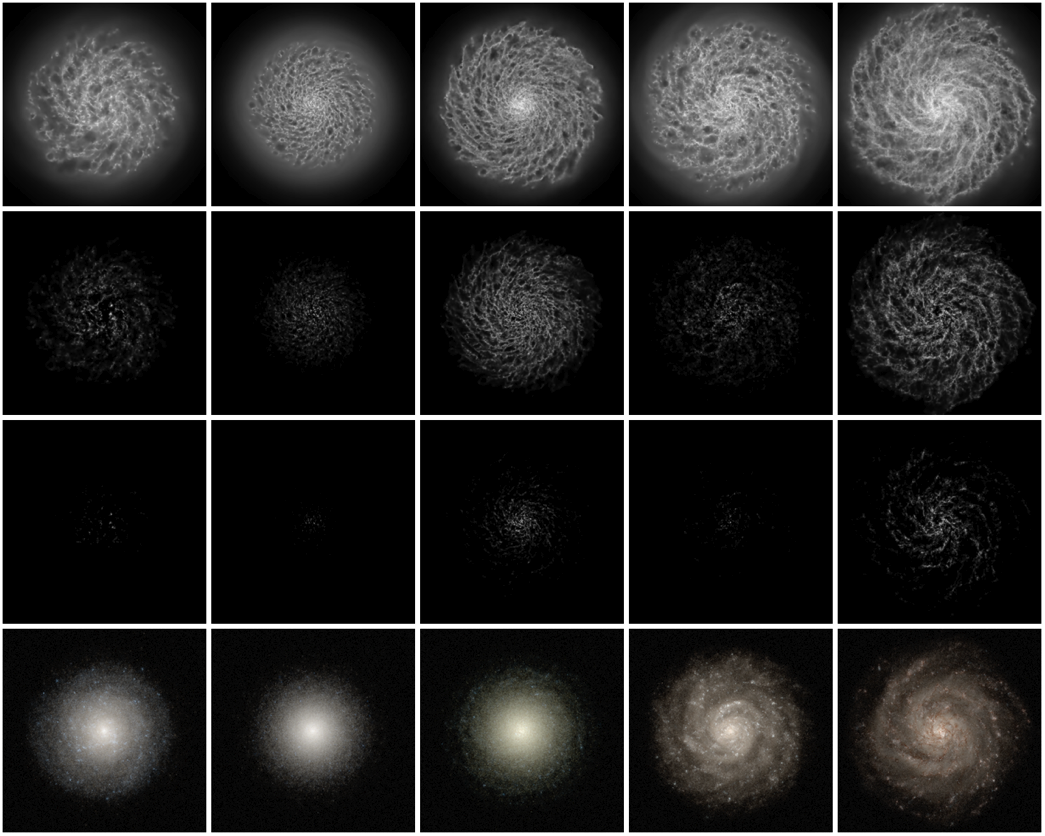}}
 \caption{Column density images from Pelupessy \& Papadopoulos
   2009, for simulations of flocculent spirals, which include the gas and
   stars. 
   The images show HI (top), H$_2$ (second line), CO (third
   line) and UBV (lowest line) from models with varying amounts of gas
 and metallicites.}
   \label{fig1}
\end{center}
\end{figure}
 \begin{figure}[t]
\begin{center}
\centerline{ \includegraphics[width=3.5in]{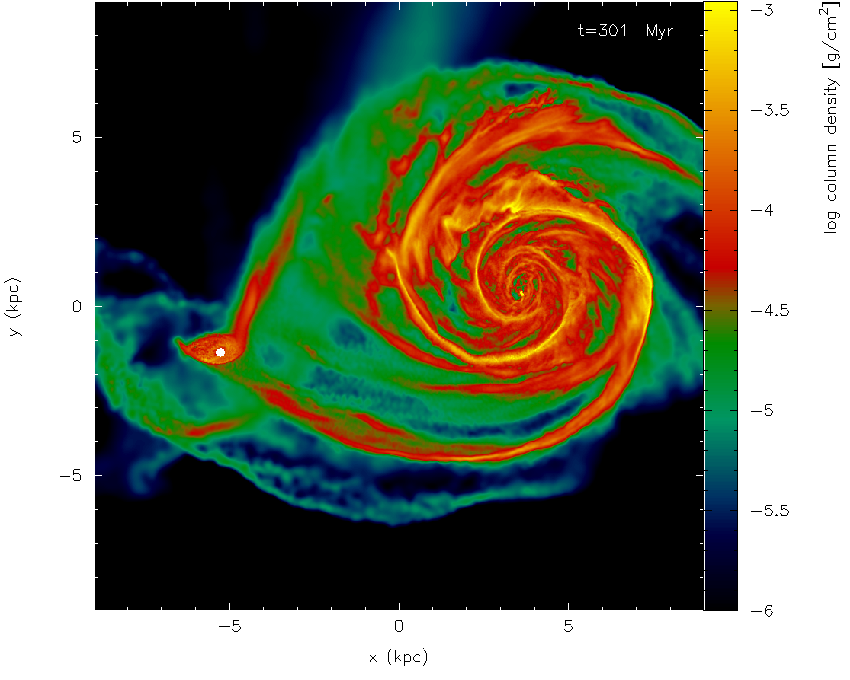}}
 \caption{This figure is taken from a calculation designed to
   reproduce the spiral structure of M51 (Dobbs et al. 2010). The
   companion galaxy, NGC 5193, is modelled as a point mass (seen as a
   white dot in the figure), and the two galaxies are given inital
   velocities and positions derived using N-body calculations (Theis
   \& Spinneker 2003).}
   \label{fig1}
\end{center}
\end{figure}
\section{Properties of molecular clouds formed in the simulations}
In Dobbs 2008, we started to determine properties of molecular
clouds formed in the simulations with a fixed spiral potential. The mass functions of the clouds have an index of
1.7-2.0, in line with observations. The cloud angular momenta also
appear in good agreement with clouds in both the Milky Way (Phillips
1999) and M33
(Rosolowsky et al. 2003). Clumps regularly undergo collisions or
interactions with other clumps, which can lead to clumps rotating in a
retrograde direction, i.e. in the opposite sense to galactic
rotation. Properties of molecular clouds have also been deteremined in
calculations which used the AMR code ENZO (Tasker \& Tan 2009), and show
remarkable similarity with SPH calculations, although there are a
number of differences between the models (e.g. presence of magnetic
field, spiral potential).

Alves (these proceedings) derive a remarkably tight mass radius relation for
molecular clouds. This very tight correlation is not apparent in the results of Heyer et al. 2009,
who re-examined the properties of clouds observed by Solomon et
al. 1987.
Numerical simulations may help explain the discrepancies in these
results, and why the clouds found by Heyer et al. 2009 tend to be
slightly unbound.

\section{Conclusion}
SPH calculations allow the ISM and molecular clouds to be studied on
galactic scales. Unlike previous particle methods, these calculations
can model individual clouds, and thus capture changes in the size,
shape, and constituent gas of the clouds. 
Hydrodynamic simulations show that when a global grand design
spiral pattern is present, GMCs form by the agglomeration of gas
clouds in the spiral shock. This process requires the gas is clumpy
(or cold), but appears independent of resolution down to at least 10 M$_{\odot}$ per
particle. It is also independent of the treatment of artificial viscosity, and
the initial particle distribution. Magnetic fields do not strongly
affect this picture, and the detailed thermodynamics do not make a
significant difference, provided there is cold gas. Self gravity can alter the large scale
distribution, both enhancing cloud
agglomeration and gravitational instabilities, the importance of self
gravity increasing with surface density.
\section{Acknowledgments}
I would like to thank Ian Bonnell and Jim Pringle for their help
throughout this research. I also thank Daniel Price for providing his
SPH code `phantom' which was used to perform the resolution test in
Section 2. The calculations reported here were performed using 
the University of Exeter's SGI Altix ICE 8200 supercomputer, and 
the HLRB-II: SGI Altix 4700 supercomputer at the Leibniz supercomputer
centre, Garching. Images included in this review 
were produced using SPLASH (Price 2007), a visualisation tool for SPH that is
publicly available at http://www.astro.ex.ac.uk/people/dprice/splash.

\end{document}